\documentclass{article}

\usepackage{graphicx}% Include figure files

\long\def\symbolfootnote[#1]#2{\begingroup%
\def\thefootnote{\fnsymbol{footnote}}\footnote[#1]{#2}\endgroup}

\begin{document}
\begin{center}

{\LARGE \bf An exactly solvable dead-layer problem}\\[10mm]

 F.\ B.\ Pedersen$^{a}$ and P.\ C.\ Hemmer$^b$\symbolfootnote[1]{Corresponding author. Email:
chr-hem@online.no\vspace{6pt}}

{\small

$^a$Det Norske Veritas, Veritasveien 1, N-1322 H\o vik, Norway

$^b$Department of Physics, Norwegian University of Science and
Technology, N-7491 Trondheim, Norway }
\end{center}

\noindent {\small A molecule consisting of two particles
interacting with a delta-function attraction, and confined to a
one-dimensional volume, is studied. From the exact solution of the
system we deduce that the center-of-mass effectively moves in a
volume reduced by an inaccessible dead layer near each wall. For a
very large volume the dead-layer width equals the size of the
molecule, defined as the expectation value of the interparticle
distance for the unconfined molecule. For finite volumes the
dead-layer width is smaller than this. For unequal particle
masses,   perturbation in the mass difference up to third order
yields the same result, and we conjecture that for this system the
dead-layer width is independent of the mass values.}

\bigskip


\section{Introduction}

For the center of a small classical ball in a container, there will be an inaccessible volume near the wall. Should the ball be incompressible, the width of the inaccessible volume is equal to the ball radius. A similar situation is expected for a bound quantum-mechanical system, a molecule or an exciton, say, in a large finite volume.
The center of mass cannot easily reach the boundary since the separate particles have to stay inside the volume. Therefore the center of mass effectively moves in a volume that compared with the actual volume is reduced by a {\it dead layer} at the boundary.
Excitons or other bound systems in quantum dots or quantum wells are prime examples (Fig.\ 1). \\

\begin{figure}[h]
\includegraphics*[width=240pt,height=430pt,angle=-90]{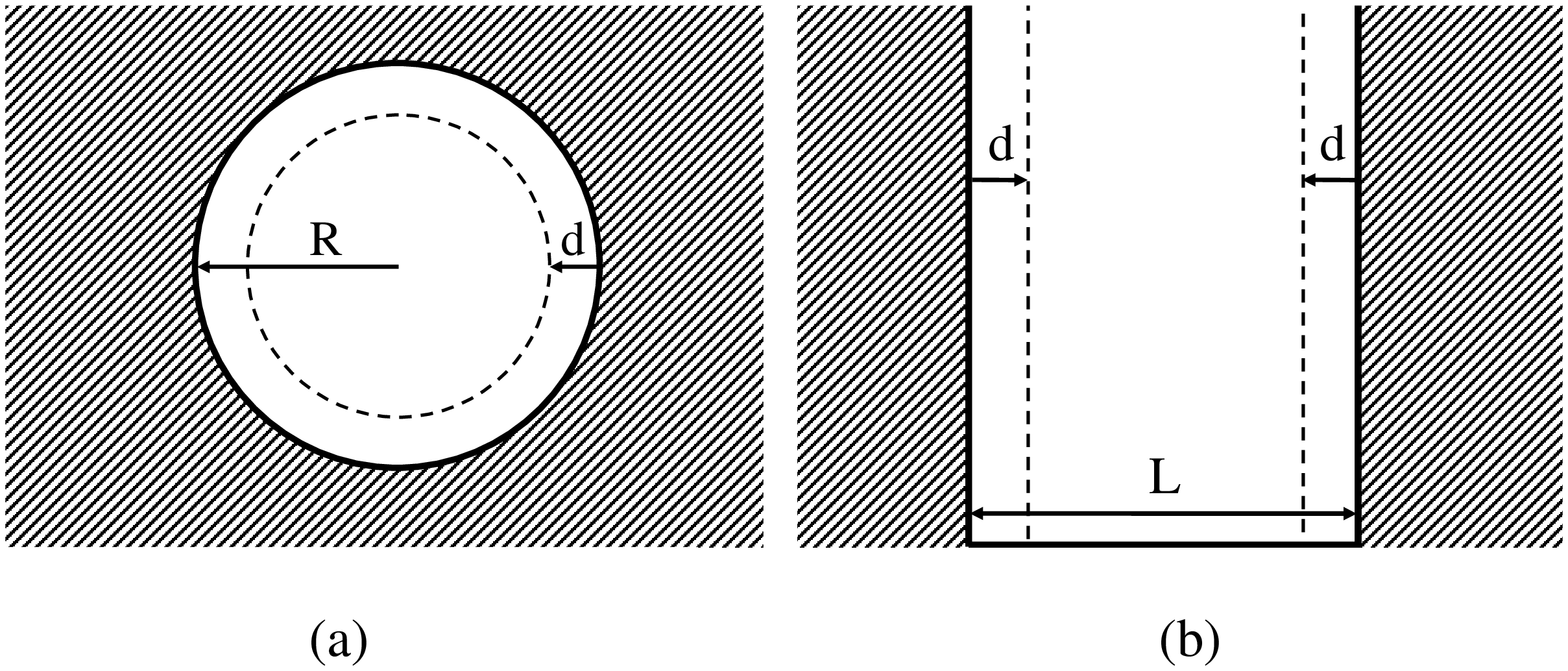}
\caption{Dead layer in a spherical quantum dot of radius $R$, and
in a quantum well of width $L$.}
\end{figure}

Consequently one expects the ground-state energy of the confined system to be of the form
\begin{equation}     E = - B +\frac{\hbar^2\pi^2}{2M(R-d)^2} \label{dot}\end{equation}
for a large spherical volume of radius $R$, and
\begin{equation}     E = - B + \frac{\hbar^2\pi^2}{2M(L-2d)^2}\label{well}\end{equation}
for a quantum well of width $L$. Here $B$ is the binding energy of
the system when unconfined, $M$ is the total mass of the system,
and $d$ measures the width of the dead layer. The notion of a dead
layer was apparently first used by Pekar \cite{Pekar}. The energy
expressions (\ref{dot}) or (\ref{well}) are commonly used by
experimentalists to fit their data \cite{NN,MM,Combescot}.

From a fundamental point of view  the dead-layer concept raises several questions:
\begin{itemize}
\item A proper dead layer should be independent of the volume in
the limit of a very large volume. While one may always write the
energy in the form (\ref{dot}) or (\ref{well}), one should show
that the value of  the dead layer parameter $d$ approaches a
limiting value when the volume increases. \item It would be
interesting to see  whether $d$ approaches its limiting value from
above or from below. Some experimental results \cite{MM} indicate
that $d$ is smaller than its limiting value when the volume is
finite. \item The size of an unconfined system of two particles
with masses $m_1$ and $m_2$ is determined by the reduced mass $\mu
=m_1m_2/(m_1+m_2)$. However, there is no reason to expect that the
dead-layer width depends on the masses merely through $\mu$.
Whether, for a given value of the reduced mass, the width of the
dead layer increases or decreases with increasing difference
between the two masses is controversial (see \cite{Combescot} and
references therein).
\end{itemize}

We will study these questions in a confined two-particle model. In
order to have an exactly solvable system we work in one dimension,
and with an attractive delta-function interaction between the
particles. For equal masses we will be able to find the {\em
exact} ground state of the confined
 system. This is of interest in itself as a new non-trivial solution of the Schr\"{o}dinger equation. In Sec.\ 2 this exact solution is worked out. The effect of nonequal masses is studied in Sec.\ 3, and our results are summerized in Sec.\ 4.

\section{The model and its exact solution}

We consider two particles with  delta-function attraction in a
one-dimensional quantum well of extension $L$ with infinite
offsets. The Schr\"{o}dinger equation is
\begin{equation}
-\frac{\hbar^2}{2m_1}\frac{\partial^2\psi}{\partial x_1^2}-\frac{\hbar^2}{2m_2}
\frac{\partial^2\psi}{\partial x_2^2} - \frac{\hbar^2}{\mu a}\;\delta(x_2-x_1)\psi = E \psi.
\label{Sch}
\end{equation}
Here  $\mu = m_1m_2/(m_1+m_2)$ is the reduced mass, and $a$ is a measure of the potential
strength. The particles live on the interval $(-L/2,L/2)$. Hence the boundary conditions imposed on the wave function $\psi(x_1,x_2)$ by the rigid walls are
\begin{equation}
\psi(x_1,\pm L/2) = \psi(\pm L/2,x_2) = 0.
\end{equation}

The Schr\"{o}dinger equation (\ref{Sch}) can alternatively be expressed in terms of
the relative coordinate $x=x_2-x_1$ and the center-of-mass coordinate $X=(m_1x_1+m_2x_2)/(m_1+m_2)$ as
\begin{equation}
-\frac{\hbar^2}{2M}\frac{\partial^2\psi}{\partial X^2}-\frac{\hbar^2}{2\mu}\frac{\partial^2\psi}{\partial x^2} - \frac{\hbar^2}{\mu a}\;\delta(x)\;\psi=E\psi,
\end{equation}
with the total mass $M=m_1+m_2$.

In the absence of walls the interaction between the two particles allows the bound state
\begin{equation}
\psi (x) = a^{-1/2}\;e^{-|x|/a},
\label{bound}\end{equation}
with binding energy
\begin{equation}
B = - \frac{\hbar^2}{2\mu a^2}.
\label{B}\end{equation}
The mean linear size of the bound state is of order $a$,
\begin{equation}
 \langle |x|\rangle = a/2.\label{size}
\end{equation}
The rms size is a factor $\sqrt{2}$ larger.

The important parameter of the problem is the dimensionless ratio $a/L$. For  the dead-layer problem we are interested in
situations in which the molecular size $a$ is much smaller than the container size $L$.

\subsection{The exact solution for equal masses}

We now assume $m_2=m_1=m$. Since the Schr\"{o}dinger equation is then symmetric in the position variables $x_1$ and $x_2$, the ground state must also possess this symmetry.

We first solve the Schr\"{o}dinger equation in the two domains $x_1<x_2$ and $x_1>x_2$, in which the pair potential does not appear explicitly. The two solutions are then matched to ensure that the wave function along the diagonal $x_1=x_2$ is continuous and has the singularity appropriate for the delta-function interaction.

In the domain $x_1>x_2$, the appropriate boundary conditions are $\psi(L/2,x_2)=0$ and $\psi(x_1,-L/2)=0$. The Schr\"{o}dinger equation (\ref{Sch}) may be solved by separation of variables. Product solutions satisfying the boundary conditions are
\begin{equation}
\psi_{\lambda_1,\lambda_2}(x_1,x_2) \propto \sinh[\lambda_1(x_1-L/2)/L]\;\sinh[\lambda_2(x_2+L/2)/L].
\label{1}\end{equation}
The two dimensionless separation constants $\lambda_1$ and $\lambda_2$ satisfy
\begin{equation}
E = -\frac{\hbar^2}{2mL^2}\;(\lambda_1^2+\lambda_2^2).
\label{energy}\end{equation} By symmetry we have in the other
domain $x_1<x_2$:
\begin{equation}
\psi_{\lambda_1,\lambda_2}(x_1,x_2) \propto \sinh[\lambda_1(x_2-L/2)/L]\;\sinh[\lambda_2(x_1+L/2)/L]. \label{2}
\end{equation}
By construction the wave function $\psi_{\lambda_1,\lambda_2}$ is now everywhere continuous.
We may write more compactly
\begin{eqnarray}
\psi_{\lambda_1,\lambda_2}& \propto &
\sinh[\lambda_1(x_1+x_2+|x_2-x_1|-L)/2L]\;\sinh[\lambda_2(x_1+x_2-|x_2-x_1|+L)/2L]\\
&=&\sinh[\lambda_1(2X+|x|-1)/2]\;\sinh[\lambda_2(2X-|x|+1)/2] \\
&\propto& -\cosh[(\lambda_1+\lambda_2)X+(\lambda_1-\lambda_2)(|x|-1)/2] \nonumber\\
 & &+\cosh[(\lambda_1-\lambda_2)X+\lambda_1+\lambda_2)(|x|-1)/2].
\label{ps}\end{eqnarray}
We have introduced dimensionless relative and center-of-mass coordinates,
\begin{equation}
X=(x_1+x_2)/2L \hspace{10mm} \mbox{ and } \hspace{10mm} x=
(x_2-x_1)/L,
\end{equation}
 and used that
$2\sinh(u)\sinh(v)=\cosh(u+v)-\cosh(u-v)$.

The superposition
\begin{equation}
\psi (X,x)= A\psi_{\lambda_1,\lambda_2}(X,x)+B\psi_{\lambda_2,\lambda_1}(X,x),
\label{sup}\end{equation}
with arbitrary constants $A$ and $B$,
is also a solution, since both terms correspond to the same energy (\ref{energy}).

It is clear that the separation parameters $\lambda_i$ cannot
always be real, since for sufficiently weak interaction the energy
must be positive, corresponding to imaginary $\lambda_i$. We
therefore take $\lambda_1$ and $\lambda_2$ complex. Since the
energy (\ref{energy}) is real, we select separation parameters
that are complex conjugates. Thus we put
\begin{equation}
    \lambda_1 = u + iv \hspace{10mm}\mbox{ and } \hspace{10mm}\lambda_2= u-iv,
\label{uv}\end{equation} with real $u$ and $v$. Moreover it is
clear that with $A=B$ in (\ref{sup}) the wave function will be
real, and we will show that this is the correct superposition for
the ground state. Using (\ref{ps}), (\ref{uv}) and (\ref{sup})
with $A=B$, we obtain
\begin{eqnarray}
\psi(X,x) &\propto& \Re e[\cosh(2ivX-u+u|x|)-\cosh(2uX-iv+iv|x|)]\\
&=& \cos(2vX)\cosh(u|x|-u)-\cosh(2uX)\cos(v|x|-v)
\end{eqnarray}

By construction this wave function obeys the boundary conditions,
and also the Schr\"{o}dinger equation for $x_1\neq x_2$. It
remains, however, to show that it may satisfy the Schr\"{o}dinger
equation with the correct interaction potential
$V(x)=-\delta(x)\hbar^2/\mu a$.

 For a general interaction potential $V$ the Schr\"{o}dinger equation in dimensionless relative and center-of-mass coordinates takes the form
\begin{equation}
-\frac{\hbar^2}{2ML^2} \frac{\partial^2\psi}{\partial X^2}-\frac{\hbar^2}{2\mu L^2}\frac{\partial^2\psi}{\partial x^2} + V \psi = E \psi.
\end{equation}
Thus the potential can be calculated from the wave function,
\begin{equation}
V = \frac{\hbar^2}{2\mu L^2\;\psi}\frac{\partial^2\psi}{\partial x^2} + \frac{\hbar^2}{2ML^2\psi}
\frac{\partial^2\psi}{\partial X^2} + E.
\end{equation}
Clearly the desired delta-function potential can only come from the first term containing a differentiation with
respect to the relative coordinate $x$. Using $d^2|x|/dx^2=2\delta(x)$ we obtain
\begin{equation}
V = \frac{\hbar^2}{\mu L^2}\;\frac{v\sin(v)\cosh(2uX)+u\sinh(u)\cos(2vX)}{\cos(v)\cosh(2uX)-
\cosh(u)\cos(2vX)}\;\;\delta(x).
\label{a}\end{equation}
The desired form in (\ref{Sch}),
\begin{equation}
V = -\frac{\hbar^2}{\mu a}\;\delta(x_2-x_1)= - \frac{\hbar^2}{\mu a L}\;\delta(x),
\label{b}\end{equation}
with dimensionless $x$, will be obtained when we take
\begin{equation}
v\sin(v) = -\frac{L}{a}\;\cos(v)
\end{equation}
and
\begin{equation}
u\sinh(u)=\frac{L}{a}\;\cosh(u).
\end{equation}

In conclusion, we have found the following exact ground state of the system:
\begin{equation}
\psi(X,x) = N[\cosh(u|x|-u)\cos(2vX)-\cosh(2uX)\cos(v|x|-v)],
\label{exact}\end{equation}
where the coordinates are scaled by $L$. The parameters $u$ and $v$ satisfy
\begin{equation}
u\tanh(u)=-v\tan(v)=L/a,
\label{parameters}\end{equation}
and the energy (\ref{energy}) is given by
\begin{equation}
E = \frac{\hbar^2}{mL^2}(v^2-u^2).
\label{E}\end{equation}

It is straightforward, but tedious, to determine the normalization constant $N$. The result is
\begin{equation}
N = \frac{1}{L}\sqrt{\frac{8uv}{(\sinh(2u)+2u)(\sin(2v)+2v)}}.
\label{norm}\end{equation}

\subsection{Discussion of the solution}

Both the parameters $u$ and $v$ may be taken nonnegative, since the wave function is invariant
under parameter sign changes. Since $u\tanh(u)$ is a monotonically increasing function, $u$ is determined
uniquely by (\ref{parameters}). The equation for $v$, on the other hand, has no unique solution.
Since the energy expression (\ref{E}) increases with increasing $v$, the ground state corresponds to the smallest solution for $v$. (\ref{parameters}) shows that $\tan(v)$ is negative, and we conclude that the relevant solution for $v$ lies in the interval $(\pi/2,\pi)$.

The boundaries of the $v$-interval correspond to special limiting cases. For $v=\pi$ we have
$u=0$ and
\begin{equation}
\psi = L^{-1}\left[\cos(2\pi X)+\cos(\pi|x|)\right] = \frac{2}{L}\cos(\pi x_1/L)\cos(\pi x_2/L),
\end{equation}
two uncoupled particles in the domain. This {\em strong confinement} limit is consistent both with the corresponding energy value, and with (\ref{parameters}), since $v=\pi$ corresponds to $a=\infty$, i.e., vanishing delta-function amplitude.

When $v\rightarrow \pi/2$ from above, eq.\ (\ref{parameters}) shows that $u$ becomes very large approaching the value $L/a$. This is the {\em weak confinement} limit, and   may be considered as $a$ being finite and $L\rightarrow \infty$.
The interesting factor in the wave function (\ref{exact}) is
\begin{eqnarray}
\cosh(u|x|-u)& \simeq &\cosh(u)\cosh(u|x|)-\sinh(u)\sinh(u|x|) \simeq \cosh(u)[\cosh(u|x|)-\sinh(u|x|)]\nonumber\\
&= &
\cosh(u) \;e^{-u|x|} \simeq \cosh(u) e^{-|x_2-x_1|/a}.
\end{eqnarray}
Using also that the normalization factor is approximately given by $1/N\simeq \sqrt{aL/2}\cosh(u)$ in this limiting case, we obtain altogether
\begin{equation}
\psi \simeq a^{-1/2}e^{-|x_2-x_1|/a}\;\;(2/L)^{-1/2}\;\cos(\pi (x_1+x_2)/L),
\end{equation}
a product of the bound state and the center-of-mass ground state in a large volume.

To illustrate the wave function inbetween these limiting cases we show in Fig.\ 2(a) $\psi(x,0)$ for different ratios $a/L$.  The cusp at $x=0$ becomes more and more pronounced with increasing  $L/a$, as expected. The second derivative near the cusp is positive only for $a<0.431 L$. In Fig.\ 2(b) $\psi(x_1,x_2)$ is shown for $L=5a$. We note that the wave function is everywhere non-negative, as must be in the ground state.\\

\begin{figure}
\includegraphics[width=180pt,height=400pt,angle=-90]{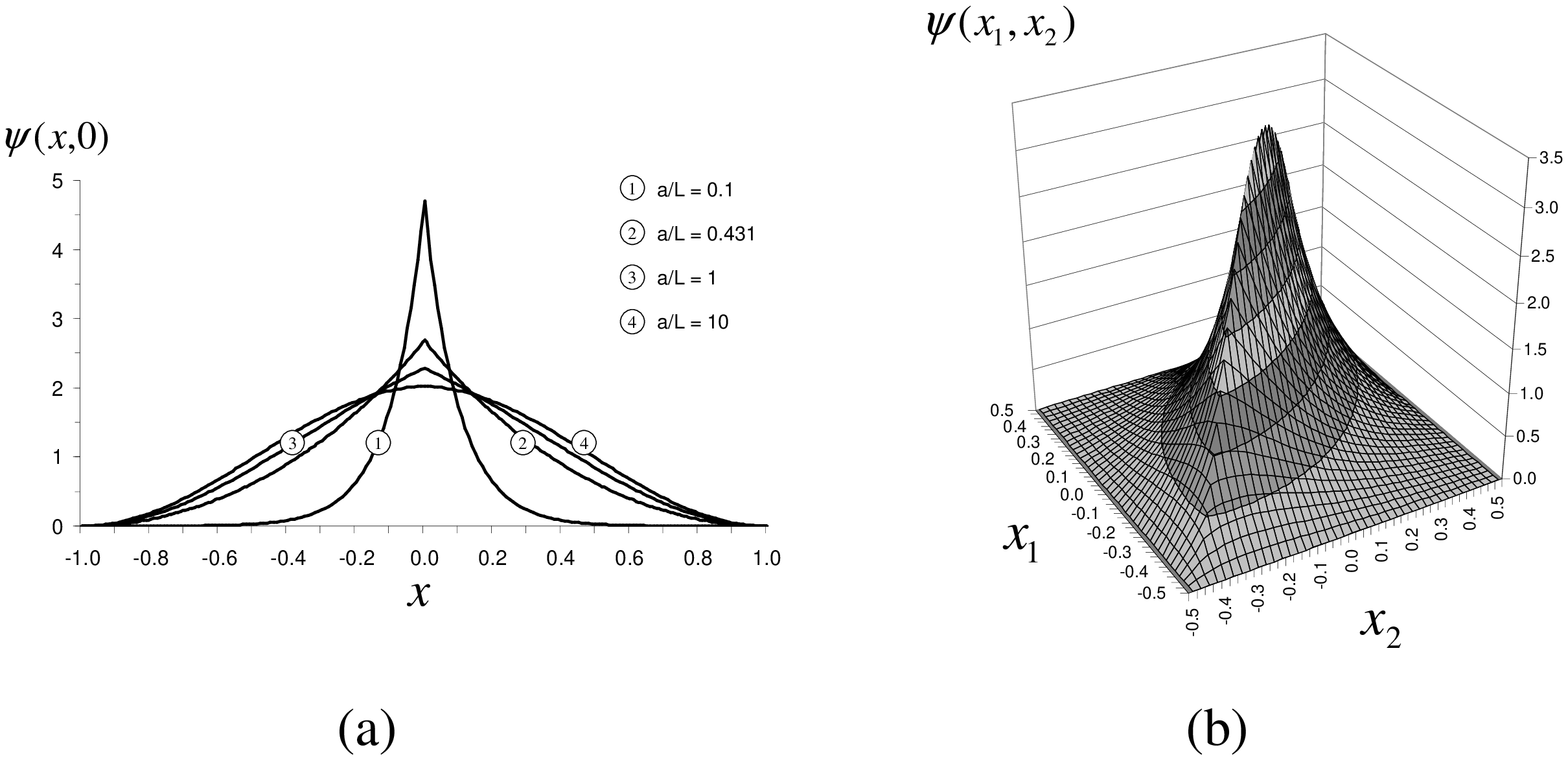}
\caption{(a) The wave function $\psi(x,X=0)$ for different values
of $a/L$. (b) The ground state $\psi(x_1,x_2)$ for $L=5a$.
Dimensionless coordinates are used.}
\end{figure}

\subsection{The dead layer}

We now investigate the energy (\ref{E}) for large volumes. For $L/a$ large we see from
(\ref{parameters}) that $u$ is large and approximately equal to $L/a$. We take $u=L/a$, since the correction terms to this value are exponentially small and can safely be ignored. Thus we have to discuss
\begin{equation}
E= -\frac{\hbar^2}{ma^2}+\frac{\hbar^2}{mL^2}\;v^2.
\label{ev}\end{equation} By eq.\ (\ref{B}) the first term is the
molecular binding energy $-B$. Writing the energy (\ref{ev}) in
the suggestive form
\begin{equation}
E = -B + \frac{\hbar^2\pi^2}{2M(L-2d)^2},
\end{equation}
we have
\begin{equation}
d = \frac{L}{2}-\frac{\pi L}{4v}.
\label{d}\end{equation}
Here $v \in (\pi/2,\pi)$ is given by $v\tan(v)=-L/a$, eq.\ (\ref{parameters}).
It is straightforward to solve this equation for $v$ perturbatively for small
$a/L$, since $v$ is very close to $\pi/2$ in this limit. We find
\begin{equation}
v = \frac{\pi}{2} \left[1+\frac{a}{L}+\frac{a^2}{L^2}+\left(1-\frac{\pi^2}{12}\right)\;\frac{a^3}{L^3}
+{\cal O}\left(\frac{a^4}{L^4}\right)\right].
\label{v} \end{equation}
Insertion into eq.(\ref{d}) gives the corresponding expansion for $d$:
\begin{equation}
d = \frac{a}{2} -\frac{\pi^2}{24}\frac{a^3}{L^2} + \ldots
\end{equation}

Thus in the limit of a large container a dead layer of width
\begin{equation}
d= a/2
\end{equation}
exists. Hence the dead-layer width equals precisely the mean
linear size (\ref{size}) of the unconfined molecule.

Moreover, when the container is finite, the effective dead layer is somewhat
reduced in size. This can be see in more in more detail by calculating $d$
 for all values of $a/L$. We must then use the full equation (\ref{parameters}) for $u$, not the weak confinement approximation $u\simeq L/a$. Fig.\ 3 shows that $d$ is a monotonically increasing function of $L/a$.\\

\begin{figure}
\includegraphics[width=240pt,height=400pt,angle=-90]{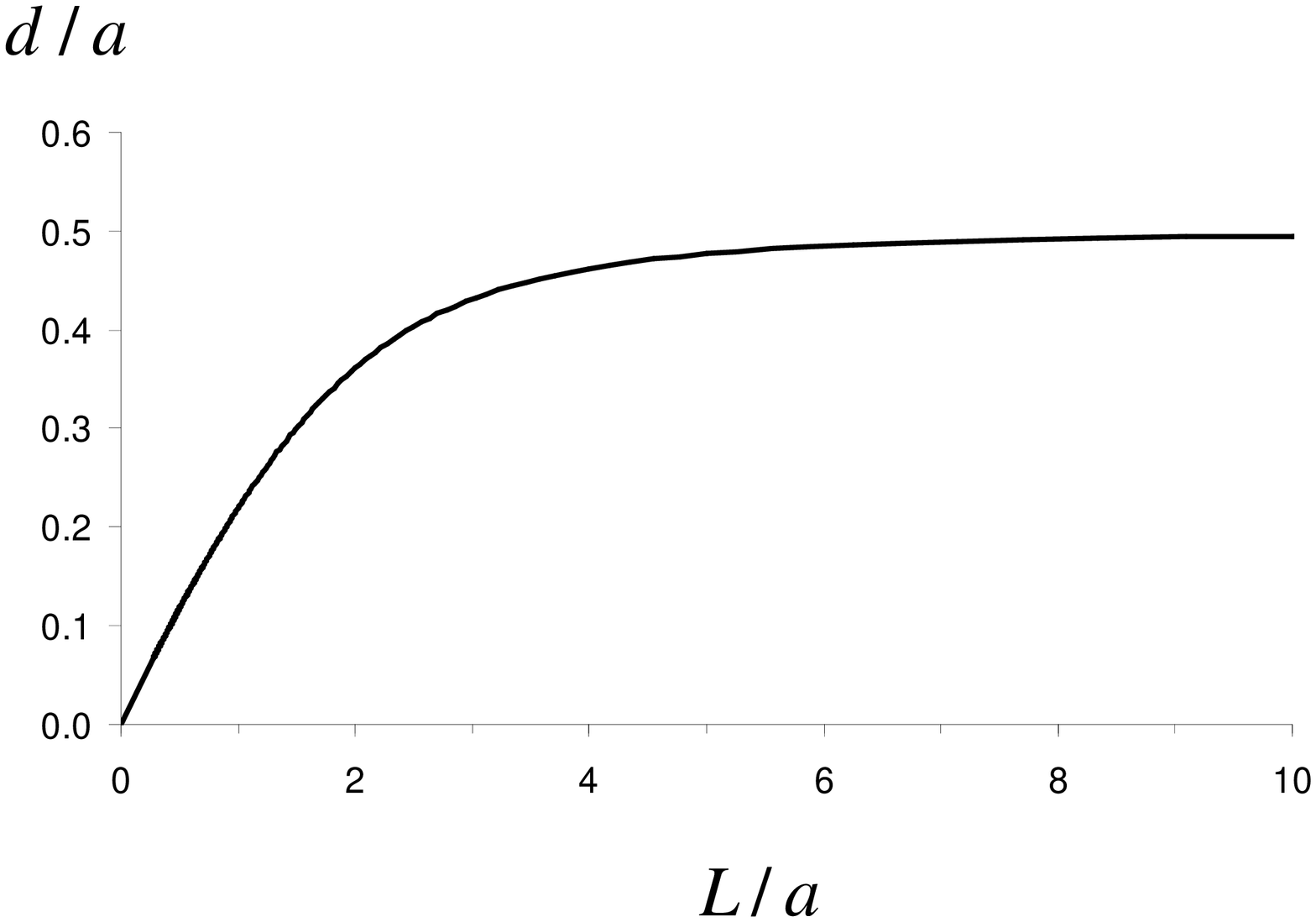}
\caption{The effective dead layer as function of $L/a$.}
\end{figure}

The molecule is expected to be deformed close to a wall. To illustrate this, we calulate  the linear size of the molecule,
\begin{equation}
\langle |x|\rangle = \frac{\int |x|\;|\psi(X,x)|^2\;dx}{\int
|\psi(X,x)|^2\;dx}, \label{size2}\end{equation} for $X$ positive,
and study how the size
 changes when the center of mass approaches the wall at $X=L/2$.
It is straightforward to show that for large $L/a$, the molecular size depends merely upon
$L/2-X$, the center-of-mass distance from the boundary, independent of the container size. The integration in (\ref{size}) yields for large $L/a$:
\begin{eqnarray}
\langle |x|\rangle & =&
a\;\frac{ \int_0^{2z}[(y-1)e^{-2z}-(2z-1)e^{-y}]^2\;y\;dy}{\int_0^{2z}[(y-1)e^{-2z}-(2z-1)e^{-y}]^2\;dy }\\
 &=&
\frac{a}{2}\;\frac{(2z-1)^2- (16z-8)e^{-2z}+(16z^4+\frac{80}{3}z^3+20z^2-9)e^{-4z}}{(2z-1)^2+(\frac{16}{3}z^3+4z^2-1)e^{-4z}},
\label{compr}\end{eqnarray}
where
\begin{equation}
z =(L/2-X)/a.
 \end{equation}
Details of the integration are not given.
For large $z\rightarrow \infty$, i.e. far from the wall, we recover from (\ref{compr}) the molecular size $\langle |x|\rangle=a/2$, as expected.
Fig.\ 4 shows how the molecule is compressed when its center of mass is forced to be close to a wall.
When the center of mass is at a dead-layer distance from the wall,  $X=L/2-a/2$ or $z=1/2$, the molecular size has by eq. (\ref{compr})  shrunk to one-half of its size in bulk.\\

\begin{figure}
\includegraphics[width=240pt,height=400pt,angle=-90]{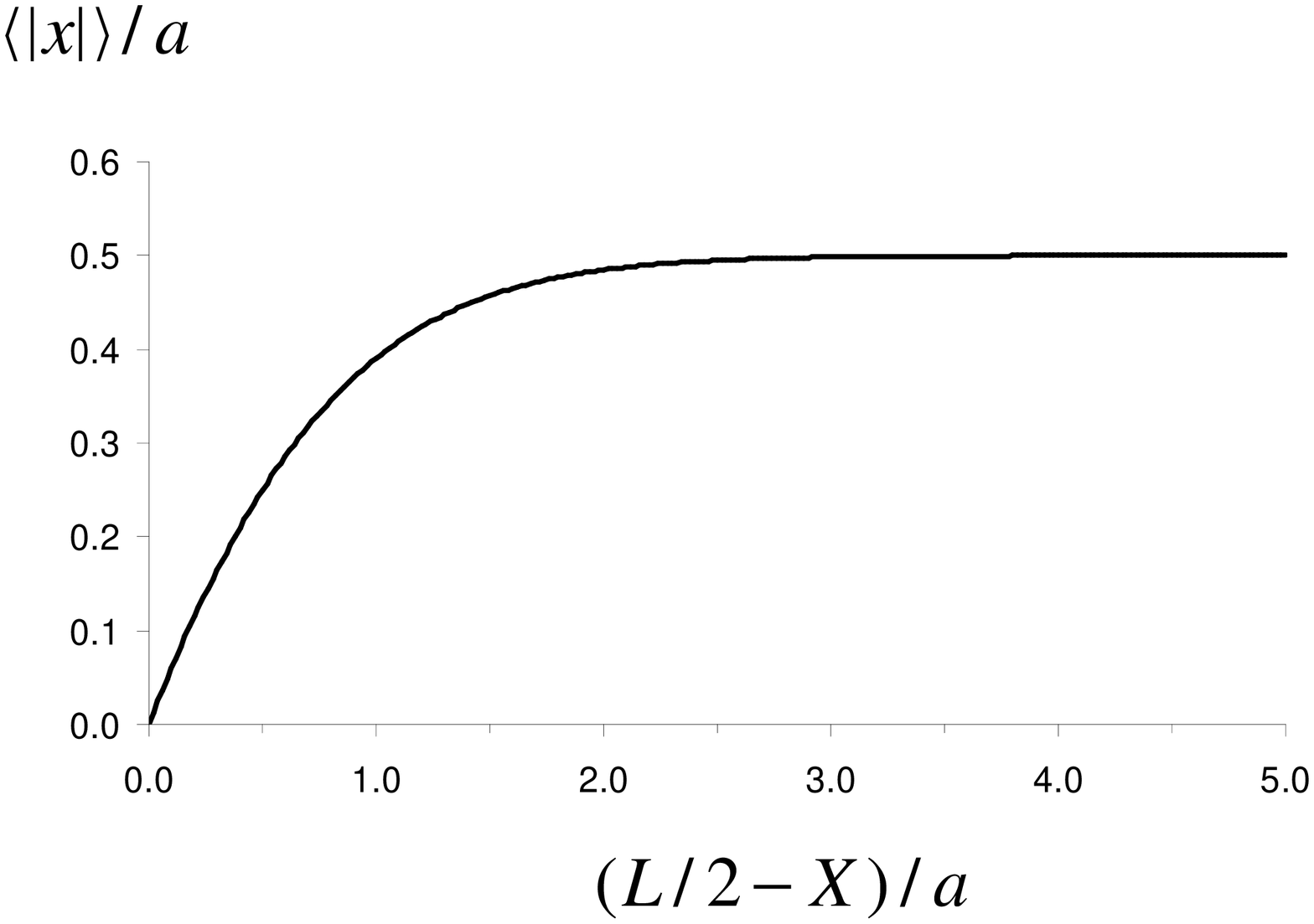}
\caption{The dimensionless molecular size $\langle |x|\rangle /a$
as function of $(L/2-X)/a$, the dimensionless distance between the
center-of-mass position and the boundary.}
\end{figure}

\section{Unequal masses}

For particles with unequal masses the hamiltonian is
\begin{equation}
H = -\frac{\hbar^2}{2m_1}\frac{\partial^2}{\partial
x_1^2}-\frac{\hbar^2}{2m_2}\frac{\partial^2}{\partial x_2^2} -
\frac{\hbar^2}{a\mu}\delta (x_2-x_1). \end{equation} The boundary
conditions are as before. In the absence of an exact solution of
the corresponding Schr\"{o}dinger equation, we now resort to
Rayleigh-Schr\"{o}dinger perturbation theory in the mass
difference.

\subsection{First-order perturbation}

 We use an unperturbed Hamiltonian
\begin{equation}
H_0 = -\frac{\hbar^2}{2m}\frac{\partial^2}{\partial
x_1^2}-\frac{\hbar^2}{2m}\frac{\partial^2}{\partial x_2^2} -
\frac{\hbar^2}{a\mu}\delta (x_2-x_1), \label{Hun}\end{equation}
with $m$ selected such that the reduced mass $\mu$ is the same for
the perturbed and the unperturbed problem:
\begin{equation}
\frac{2}{m}= \frac{1}{m_1}+\frac{1}{m_2}=\frac{1}{\mu}.
\label{mu}\end{equation}
The perturbation is then
\begin{equation}
H'= H-H_0 =
\frac{\hbar^2}{4}\left(\frac{1}{m_1}-\frac{1}{m_2}\right)\left(\frac{\partial^2}{\partial
x_1^2}-\frac{\partial^2}{\partial x_2^2}\right).
\label{Hpert}\end{equation} The ground state (\ref{exact}) of the
unperturbed problem (\ref{Hun}) is symmetric with respect to
interchanging $x_1$ and $x_2$, while the perturbation $H'$ is
unsymmetric. Hence the first-order perturbation contribution
vanishes:
\begin{equation}
E^{(1)}=\langle 0 |H'|0\rangle = 0.
\end{equation}

\subsection{Second-order perturbation}

The second-order perturbation contribution to the ground state is
\begin{equation}
E^{(2)} = \sum_{k\neq 0} \frac{|\langle k|H'|0\rangle|^2}{E_0^{0}-E_k^{0}}.
\label{second}\end{equation}
To evaluate the sum we use the Dalgarno-Lewis technique \cite{Dalgarno}. Suppose we can find an operator $F$ satisfying
\begin{equation}
H'|0\rangle = (FH_0-H_0F)|0\rangle =[F,H_0]|0\rangle.
\label{DL}
\end{equation}
Then
\begin{equation}
\langle k|H'|0\rangle = \left(E_0^0-E_k^{0}\right) \langle k|F|0\rangle,
\end{equation}
and the sum in (\ref{second}) may be written as
\begin{equation}
 E^{(2)}=\sum_{k\neq 0} \frac{|\langle k|H'|0\rangle|^2}{E_0^{0}-E_k^{0}} = \sum_{k\neq 0}\langle 0|H'|k\rangle \langle k|F|0\rangle=\langle 0|H'F|0\rangle-\langle 0|H'|0\rangle \langle 0|F|0\rangle.
\label{E2}\end{equation}
The great advantage is that the summation is avoided, and that merely the unperturbed ground state enters.

In relative and center-of-mass coordinates, we have
\begin{equation}
H_0 = -\frac{\hbar^2}{m} \frac{\partial^2}{\partial x^2}-\frac{\hbar^2}{4m}\frac{\partial^2}{\partial X^2}-\frac{\hbar^2}{a\mu} \;\delta (x),
\label{H0rel}\end{equation}
and
\begin{equation}
H'= \frac{\hbar^2}{2}\left(\frac{1}{m_1}-\frac{1}{m_2}\right)\;\frac{\partial}{\partial x}\frac{\partial}{\partial X}.
\end{equation}
For the problem at hand the Dalgarno-Lewis operator $F$ may be taken as
\begin{equation}
F=\frac{m}{4}\left(\frac{1}{m_1}-\frac{1}{m_2}\right) \;x\;\frac{\partial}{\partial X}.
\end{equation}
In (\ref{DL}) the commutator between $F$ and $H_0$ enters. Since $\partial/\partial X$ commutes with $H_0$, eq.(\ref{H0rel}), we have
\begin{equation}
[F,H_0] = - \frac{\hbar^2}{4}
\left(\frac{1}{m_1}-\frac{1}{m_2}\right)\;
\left[x,\frac{\partial^2}{\partial x^2}\right]
\frac{\partial}{\partial X} =
\frac{\hbar^2}{2}\left(\frac{1}{m_1}-\frac{1}{m_2}\right)
\frac{\partial}{\partial x}\frac{\partial}{\partial X}\equiv H',
\end{equation}
thus the Dalgarno-Lewis requirement (\ref{DL}) is
fulfilled \cite{DL_Boundary}.

It only remains to determine the right-hand side of eq.(\ref{E2})
\begin{equation}
E^{(2)} = \frac{\hbar^2 m}{8} \left(\frac{1}{m_1}-\frac{1}{m_2}\right)^2\;\left\langle 0\left|\frac{\partial}{\partial x}\;x\;\frac{\partial^2}{\partial X^2}\right|0\ \right\rangle.
\label{e2}\end{equation}
The evaluation of the ground state average (\ref{e2}) is tedious and is referred to the appendix.
We obtain
\begin{equation}
E^{(2)} = -\frac{\hbar^2\pi^2}{16 m L^2}\left(\frac{m}{m_1}-\frac{m}{m_2}\right)^2\;\left(1+2\frac{a}{L}\right),
\label{EEE}\end{equation}
to first order in the small parameter $a/L$.

\subsection{Third-order perturbation}

The third-order perturbation contribution to the ground-state energy would depend on the masses $m_1$ and $m_2$ as follows
\begin{equation}
E^{(3)} \propto \left(\frac{1}{m_1}-\frac{1}{m_2}\right)^3.
\end{equation}
But it would be unphysical if the energy contribution  would change sign by interchanging the mass values. Thus the proportionality constant must be zero,
\begin{equation}
E^{(3)} = 0.
\end{equation}

\subsection{Result and a conjecture}

Up to third order in the mass difference we have obtained the ground state energy
\begin{equation}
E = E_0-\frac{\hbar^2\pi^2}{16 m
L^2}\left(\frac{m}{m_1}-\frac{m}{m_2}\right)^2
\left(1+2\frac{a}{L}\right)= - B + \frac{\hbar^2\pi^2}{4m
L^2}+\frac{\hbar^2\pi^2}{2mL^2}\frac{a}{L} -
\frac{\hbar^2\pi^2}{16 m
L^2}\left(\frac{m}{m_1}-\frac{m}{m_2}\right)^2
\;\left(1+2\frac{a}{L}\right), \label{res}\end{equation} to first
order in $a/L$.

With unequal masses the total mass $m_1+m_2$ of the molecule is no longer $2m$. Using the connection (\ref{mu}) between the masses, we have
\begin{equation}
\frac{1}{m_1+m_2}-\frac{1}{2m} = \frac{1}{m_1+m_2}-\frac{1}{m_1}-\frac{1}{m_2} = - \frac{m_1m_2}{4(m_1+m_2)}\left(\frac{1}{m_1}-\frac{1}{m_2}\right)^2 = - \frac{m}{8}\left(\frac{1}{m_1}-\frac{1}{m_2}\right)^2,
\end{equation}
i.e.
\begin{equation}
\frac{1}{m_1+m_2} = \frac{1}{2m}\left[1-\frac{1}{4}\left(\frac{m}{m_1}-\frac{m}{m_2}\right)^2\right]
\end{equation}
Our result (\ref{res}) takes then the simple form
\begin{equation}
E = -B + \frac{\hbar^2\pi^2}{2(m_1+m_2)L^2} \left(1+2\frac{a}{L}\right)\simeq-B +\frac{\hbar^2\pi^2}{2(m_1+m_2)(L-a)^2}
\end{equation}
to first order in $a/L$. Hence the dead-layer width is $a/2$, the
same value as for equal masses.

Thus perturbation in the mass difference gives no contribution to
the dead-layer width in first, second and third order. Although
contributions in higher-order perturbation theory cannot be
excluded a priori, we {\em conjecture} that in one dimension the
dead layer is determined solely by the reduced mass, {\em
independent} of the mass ratio.

Intuitively one may perhaps understand a null result in one dimension as follows. For unequal masses one particle will have a shorter average distance to the center of mass than the other particle. In one dimension the two distances are given by
\begin{equation}
\langle |x_1-X|\rangle = \langle |x|\rangle\;\frac{m_2}{m_1+m_2}\; \mbox{ and }\;
\langle |x_2-X|\rangle = \langle |x|\rangle\;\frac{m_1}{m_1+m_2}.
\label{int}\end{equation}
If one believe that the excluded zone is due to the short and the long distance with equal probability, the effective width of the dead layer should be determined by the average of the two distances in (\ref{int}), which  is {\em independent} of the mass ratio. This heuristic argument does not apply in higher dimensions.

\section{Concluding remarks}

We have obtained an exact solution for a system of  two interacting particles, each of mass $m$, confined to a one-dimensional volume of size $L$ between hard walls.
A dead layer is shown to exist in the limit $L\rightarrow \infty$, and the dead-layer width is precisely equal to the size $\langle |x_2-x_1|\rangle$ of the unconfined two-particle system. For finite $L$ the dead-layer width is smaller than this value. We have also determined the deformation of the system close to a wall.

When the two particles have unequal masses, we have determined the
dead-layer width by  perturbing in the mass difference. Up to
third order in the mass difference the dead-layer width is
unchanged, and we conjecture that in one dimension, and for a
given reduced mass, the dead-layer width is independent of the
mass values.

\begin{center}
{\Large \bf Appendix}
\end{center}

The second-order perturbation contribution (\ref{e2}) to the energy is
\begin{equation}
E^{(2)} =
\frac{\hbar^2}{8m}\left(\frac{m}{m_1}-\frac{m}{m_2}\right)^2\;G,
\label{ee} \end{equation} with
\begin{equation}
G=\left\langle 0\left|\frac{\partial}{\partial x}\;x\;\frac{\partial^2}{\partial X^2}\right|0\ \right\rangle
\label{G}.\end{equation}
Now we evaluate this contribution, using
the ground state (\ref{exact}),
\begin{equation}
\langle X,x|0\rangle= \psi(X,x) = N[\cosh(u|x|-u)\cos(2vX)-\cosh(2uX)\cos(v|x|-v)].
\label{exact1}\end{equation}

Since both the operator in (\ref{G}) and the wave function (\ref{exact1}) have even parity in $x$ and in $X$, we may restrict ourselves to positive $X$ and $x$, and multiply the result by 4:
\begin{equation}
G = 4 \int_0^1 dx \int_0^{(1-x)/2} dX\; \psi(X,x) \frac{\partial}{\partial x}\;x\;\frac{\partial^2}{\partial X^2}\;\psi(X,x)
\label{G1}\end{equation}

A further simplification is that we need the result only to zeroth and first order in the small parameter $a/L$. We use that the parameter $u$ is very large, $u \simeq L/a$ to dominating order, and that $v$ has the expansion (\ref{v}),
\begin{equation}
v = \frac{\pi}{2}\left[1+\frac{a}{L} + \frac{a^2}{L^2} + {\cal O}\left(\frac{a^3}{L^3}\right)\right].
\label{vv}\end{equation}
The dominating behavior of the normalization constant (\ref{norm}) is as follows,
\begin{equation}
N^2=\frac{1}{L^2}\;\frac{8uv}{(\sinh(2u)+2u)(\sin(2v)+2v)} =
\frac{8}{L^2}u e^{-2u} \left(1+\frac{a}{L}\right).\label{NN}
\end{equation}
In the wave function (\ref{exact1}) we use that
\begin{equation}
\cosh(ux-u)= \cosh(u)[\cosh(ux)-\tanh(u)\sinh(ux)]\simeq \cosh(u)[\cosh(ux)-\sinh(ux)]\simeq {\textstyle \frac{1}{2}}e^{u} e^{-ux},
\label{f}\end{equation}
since $\tanh (u) = 1$ plus exponentially small terms. The term (\ref{f}) contains the factor $e^u$ and it will dominate the second  term in $\psi$, eq.(\ref{exact1}). Although all integrations may be performed exactly, it suffices for our purpose to approximate the wave function (\ref{exact1}) as follows
\begin{equation}
\psi = {\textstyle \frac{1}{2}}N e^{uy}\;\cos(2vX),
\end{equation}
where $y=1-x$. Thus
\begin{eqnarray}
G &=&- N^2 v^2 \int_0^1 dx \; e^{2u(1-x)}(1-ux)\int_0^{(1-x)/2}  \cos^2(2vX)\;dX \nonumber\\
& =&
-N^2 v^2 \int_0^1  e^{2uy}(1-u+uy) \left[\frac{y}{4}+\frac{\sin(2vy)}{8v}\right] dy\nonumber\\
&=& - N^2v^2 e^{2u} \left[\frac{1}{16u}+\frac{u^3\sin(2v)+3uv^2\sin(2v)-2v^3\cos(2v)}{32v(u^2+v^2)^2}\right]\simeq -N^2v^2 e^{2u} \left[\frac{1}{16u}+\frac{\sin(2v)}{32uv}\right],\nonumber
\end{eqnarray}
using that $u=L/a$ plus exponentially small terms. By means of the expansion (\ref{vv}) for $v$ and the expression (\ref{NN}) for the normalization constant $N$, we obtain
to the order of interest
\begin{equation}
G=-N^2\pi^2\frac{1}{16u}e^{2u}\left(1+\frac{a}{L}\right) =-\frac{\pi^2}{2L^2} \left(1+2\frac{a}{L}\right).
\end{equation}
By insertion into (\ref{ee}) we obtain
\begin{equation}
E^{(2)} = -\frac{\hbar^2\pi^2}{16mL^2}\left(\frac{m}{m_1}-\frac{m}{m_2}\right)^2\;\left(1+2\frac{a}{L}\right).
\end{equation}
This is eq.(\ref{EEE}) in the main text.


\begin{thebibliography}{99}
\bibitem{DL_Boundary} By construction the perturbation expansion
involves only functions fullfilling the boundary conditions.
Confinement potentials can therefore be omitted from the
Hamiltonian and hence the commutator relations.
\end{thebibliography}
\end{document}